\documentclass[preprint,12pt]{elsarticle}

\usepackage{graphicx,subfigure}
\usepackage{amssymb}
\usepackage{amsmath}
\usepackage{bm}
\usepackage{lineno} 
\usepackage{natbib}
\usepackage[squaren]{SIunits}

\begin{document}

\begin{frontmatter}

\title{Mitigation of numerical Cerenkov radiation and instability using a hybrid finite difference-FFT Maxwell solver and a local charge conserving current deposit}
 
\author[UCLAEE]{Peicheng Yu}
\ead{tpc02@ucla.edu}
\author[THUACC]{Xinlu Xu} 
\author[UCLAPH]{Adam Tableman}
\author[UCLAPH]{Viktor K. Decyk}
\author[UCLAPH]{Frank S. Tsung}
\author[LLNL]{Frederico Fiuza}
\author[UCLAPH]{Asher Davidson}
\author[IST]{Jorge Vieira}
\author[IST,ISCTE]{Ricardo A. Fonseca}
\author[THUACC]{Wei Lu}
\author[IST]{Luis O. Silva}
\author[UCLAEE,UCLAPH]{Warren B. Mori}

\address[UCLAEE]{Department of Electrical Engineering, University of California Los Angeles, Los Angeles, CA 90095, USA}
\address[THUACC]{Department of Engineering Physics, Tsinghua University, Beijing 100084, China}
\address[UCLAPH]{Department of Physics and Astronomy, University of California Los Angeles, Los Angeles, CA 90095, USA}
\address[LLNL]{Lawrence Livermore National Laboratory, Livermore, California, USA}
\address[IST]{GOLP/Instituto de Plasma e Fus\~ao Nuclear, Instituto Superior T\'ecnico, Universidade de Lisboa, Lisbon, Portugal}
\address[ISCTE]{ISCTE - Instituto Universit\'ario de Lisboa, 1649--026, Lisbon, Portugal}

\begin{abstract}
A hybrid Maxwell solver for fully relativistic and electromagnetic (EM) particle-in-cell (PIC) codes is described. In this solver, the EM fields are solved in $k$ space by performing an FFT in one direction, while using finite difference operators in the other direction(s). This solver eliminates the numerical Cerenkov radiation for particles moving in the preferred direction. Moreover, the numerical Cerenkov instability (NCI) induced by the relativistically drifting plasma and beam can be eliminated using this hybrid solver by applying strategies that are similar to those recently developed for pure FFT solvers. A current correction is applied for the charge conserving current deposit to correctly account for the EM calculation in hybrid Yee-FFT solver. A theoretical analysis of the dispersion properties in vacuum and in a drifting plasma for the hybrid solver is presented, and compared with PIC simulations with good agreement obtained. This hybrid solver is applied to both 2D and 3D Cartesian and quasi-3D (in which the fields and current are decomposed into azimuthal harmonics) geometries. Illustrative results for laser wakefield accelerator simulation in a Lorentz boosted frame using the hybrid solver in the 2D Cartesian geometry are presented, and compared against results from 2D UPIC-EMMA simulation which uses a pure spectral Maxwell solver, and from OSIRIS 2D lab frame simulation using the standard Yee solver. Very good agreement is obtained which demonstrates the feasibility of using the hybrid solver for high fidelity simulation of relativistically drifting plasma with no evidence of the numerical Cerenkov instability.
\end{abstract}


\end{frontmatter}


\section{Introduction}
\label{sect:intro}
Fully relativistic, electromagnetic particle-in-cell (PIC) codes are widely used to study a variety of plasma physics problems. In many cases the solver for Maxwell's equations in PIC codes use the finite-difference-time-domain (FDTD) approach where the corresponding differential operators are local. This locality leads to advantages in parallel scalability and ease in implementing boundary conditions. However, when using PIC codes to model physics problems, including plasma based acceleration \cite{LWFA} in the Lorentz boosted frame,  relativistic collisionless shocks \cite{shock1,shock2}, and fast ignition \cite{wilks, tonge,may} particles or plasmas stream across the grid with speeds approaching the speed of light. In these scenarios, the second order FDTD Maxwell solvers support light waves with phase velocities less than the speed of light. This property of the FDTD solver leads to numerical Cerenkov radiation from a single particle that is moving near the speed of light. In addition, when beams or plasmas are moving near the speed of light across the grid a violent numerical instability known as the numerical Cerenkov instability (NCI) arises due to the unphysical coupling of electromagnetic modes and the  Langmuir modes (main and higher order aliased beam resonance). The beam resonances are at $\omega + 2\pi\mu/\Delta t=(k_1+ 2\pi\nu_1/\Delta x_1)v_0$, where $\mu$ and $\nu_1$ refer to the time and space aliases, $\Delta t$ and $\Delta x_1$ are the time step and grid size,  and the plasma is drifting relativistically at a speed $v_0$ in the $\hat{1}$-direction.

The NCI was first studied more than 40 years ago \cite{Godfrey1974}. However, it has received much renewed attention \cite{GodfreyJCP2013,XuCPC13, YuarXiv14, YuJCP14,GodfreyPSATD, GodfreyFDTDnotes, GodfreyIEEE2014} since the identification \cite{Nagata,YuAAC12} of this numerical instability as the limiting factor for carrying out relativistic collisionless shock simulations \cite{shock1,shock2}, and Lorentz boosted frame simulations \cite{Vay2007PRL,Martins2010NatPhys,VayJCP2011,MartinsCPC} of laser wakefield acceleration (LWFA) \cite{LWFA}.

This early and recent work on the NCI \cite{Godfrey1974,GodfreyJCP2013,XuCPC13,YuarXiv14,YuJCP14,YuAAC12,Godfrey1975} have shown that the NCI inevitably arises in EM-PIC simulations when a plasma (neutral or non-neutral) drifts across a simulation grid with a speed near the speed of light. Analysis shows that it is due to the unphysical coupling of electromagnetic (EM) modes and Langmuir modes (including those due to aliasing). As a result, significant recent effort has been devoted to the investigation and elimination of the NCI so that high fidelity relativistic plasma drift simulations can be routinely performed \cite{YuarXiv14,YuJCP14,GodfreyPSATD,GodfreyFDTDnotes,VayJCP2011,MartinsCPC}.  

In previous work Ref. \cite{XuCPC13,YuarXiv14}, we examined the NCI properties for the second order Yee solver \cite{Yee}, as well as a spectral solver \cite{dawsonrmp,lin} (in which Maxwell's equation are solved in multi-dimensional $\vec{k}$ space). We note that what we refer to as simply a spectral solver, others \cite{VayJCP2013} refer to as a pseudo-spectral time domain (PSTD) solver. The NCI theory developed in \cite{XuCPC13,YuarXiv14} were general and it could be applied to any Maxwell solver. It was found that in the simulation parameter space of interest the fastest growing NCI modes of these two solvers are the $(\mu,\nu_1)=(0,\pm 1)$ modes, where $\mu$ and $\nu_1$ defined above are the temporal aliasing, and spatial aliasing in the drifting direction of the plasma. The $(\mu,\nu_1)=(0,\pm 1)$ modes for both solvers reside near the edge of the fundamental Brillouin zone (for square or cubic cells), and can be eliminated by applying a low-pass filter. However, due to the subluminal EM dispersion along the direction of the drifting plasma in the Yee solver, the main NCI mode $(\mu,\nu_1)=(0,0)$ of the Yee solver has a growth rate that is of the same order as its $(\mu,\nu_1)=(0,\pm 1)$ counterpart, and these modes reside close to the modes of physical interest. However, the $(\mu,\nu_1)=(0,0)$ NCI mode in the spectral solver has a growth rate one order of magnitude smaller than the $(\mu,\nu_1)=(0,\pm 1)$ modes due to the superluminal dispersion of spectral (FFT based) solver. Furthermore, as shown in \cite{YuarXiv14} these $(\mu,\nu_1)=(0,0)$ modes can be moved farther away from the physics modes  and their harmonics  by reducing the time step in the spectral solver, and can be fully eliminated by slightly modifying the EM dispersion in the spectral solver. Using these methods, it was demonstrated in \cite{YuarXiv14} that a spectral EM-PIC can perform high fidelity simulations of relativistically drifting plasmas where the LWFA physics is highly nonlinear with no evidence of the NCI.

A multi-dimensional spectral Maxwell solver has a superluminal dispersion relation in all the propagation directions. This is due to the fact that the first order spatial derivatives in the Maxwell's equation are greater than $N$-th order accurate (where $N$ is the number of grids) since we are solving the Maxwell's equation in $\vec{k}$ space. This superluminal dispersion relation leads to highly localized  $(\mu,\nu_1)=(0,0)$ NCI modes and the reduction of their growth rates (compared with their Yee solver counterpart). In this paper, we propose to use a hybrid Yee-FFT solver, in which the FFT is performed in only one direction, namely the drifting direction of the plasma, while keeping the finite difference form of the Yee solver in the directions transverse to the drifting direction. In other words,  EM waves moving in the $\hat 1$ direction will have a superluminal dispersion (due to the $N$-th order accurate spatial derivatives) while those moving in the  $\hat 2$ (and $\hat 3$ in 3D) directions will have a subluminal dispersion due to the second-order-accurate spatial derivatives. The advantages of this approach over a full FFT solver is that the field solver is local in the transverse directions so that better parallel scalability than a fully FFT based solver can be achieved (assuming the same parallel FFT routines are used). In addition, it is easier to include a single FFT into the structure of mature codes such as OSIRIS \cite{osiris}. Furthermore, this idea works well with a quasi-3D algorithm that is PIC in $r-z$ and gridless in $\phi$ \cite{quasi3d,davidson}, where the FFTs cannot be applied in the $\hat r$ direction. We note that recently a method for achieving improved scalability for FFT based solvers was proposed \cite{VayJCP2013} in which FFTs are used within each local domain, but it introduces as yet unquantified errors in the longitudinal fields. The relative advantages and tradeoffs between the variety of approaches being proposed will be better understood as they begin to be used on real physics problems.

We use the theoretical framework for the NCI developed in Ref. \cite{XuCPC13,YuarXiv14} to study the NCI of the hybrid solver. As we show below, the fastest growing NCI modes for the proposed hybrid solver behave similarly to those for the spectral solver. In $\vec k$ space they reside at the edge of the fundamental Brillouin zone for square or cubic cells. More importantly, the $(\mu,\nu_1)=(0,0)$ NCI mode for the hybrid solver has almost the same properties (pattern, growth rates) as that of a spectral solver. The NCI can therefore be efficiently eliminated in the  hybrid solver by applying the same strategy as in the spectral solver. Moreover,  simulations have shown that the NCI properties of the quasi-3D $r-z$ PIC and gridless in $\phi$ algorithm \cite{quasi3d,davidson} are similar to that of 2D Cartesian geometry \cite{YuAAC14}. Therefore,  the idea of a hybrid Yee-FFT solver can be readily applied to quasi-3D geometry. We also note that the use of local FFTs in domains along $z$  \cite{VayJCP2013} could be also be used within the hybrid approach described here.  

In this paper, we first discuss the algorithm for the hybrid Yee-FFT Maxwell solver in section \ref{sect:algorithm}. In section \ref{sect:nci}, we apply the theoretical framework in Ref. \cite{XuCPC13,YuarXiv14} to study the NCI properties of the hybrid solver analytically and in PIC simulations. We compare OSIRIS  \cite{osiris} results with the hybrid solver against UPIC-EMMA \cite{YuJCP14} results with a fully spectral (FFT based) solver. In section \ref{sect:nci00clean}, it is shown that the strategies used to eliminate the NCI for purely spectral solvers are also valid for the hybrid solver. In section \ref{sect:quasi3dhybrid}, we extend the hybrid solver idea to the quasi-3D algorithm in OSIRIS and present simulation studies of the NCI properties in this geometry. We then present 2D OSIRIS simulations of LWFA in a Lorentz boosted frame using the new hybrid solver. Very good agreement is found when comparing simulation results using the hybrid solver in OSIRIS against results from 2D lab frame OSIRIS using Yee solver and 2D Lorentz boosted frame UPIC-EMMA \cite{YuJCP14} simulations using spectral solver.  Last, in section \ref{sect:summary} we summarize the results and mention directions for future work.

\section{Hybrid Yee-FFT solver}
\label{sect:algorithm}
The basic idea of the hybrid Yee-FFT solver is that the theoretical framework developed in \cite{XuCPC13,YuarXiv14} indicates that the NCI is easier to eliminate when EM waves are superluminal along the direction of the plasma drift. This can be accomplished with higher order solvers or with 
an FFT based solver in  the drifting direction of the plasma (denoted as $\hat 1$-direction). We note that it is more difficult to satisfy strict charge conservation (Gauss's law) for higher order finite difference solvers. Here we replace the finite difference operator of the first spatial derivative $\partial/\partial x_1$ in the Maxwell's equation in Yee solver with its FFT counterpart that has an accuracy greater than order $N$. We then correct for this change in the current deposit to maintain strict charge conservation. Without loss of generality, in the following we will briefly describe the algorithm of the Yee-FFT solver in two-dimensional (2D) Cartesian coordinate. The straightforward extension to the 3D Cartesian case is also discussed.

\subsection{Algorithm}
We start from the standard algorithm for a 2D Yee solver, in which the electromagnetic fields $\vec E$ and $\vec B$ are advanced by solving Faraday's Law and Ampere's Law:
\begin{align}\label{eq:yee2db1}
B^{n+\frac{1}{2}}_{1,i1,i2+\frac{1}{2}}=&~B^{n-\frac{1}{2}}_{1,i1,i2+\frac{1}{2}}-c\Delta t\times \frac{E^{n}_{3,i1,i2+1}-E^{n}_{3,i1,i2}}{\Delta x_2}\\
\label{eq:yee2db2}B^{n+\frac{1}{2}}_{2,i1+\frac{1}{2},i2}=&~B^{n-\frac{1}{2}}_{2,i1+\frac{1}{2},i2}+c\Delta t\times \frac{E^{n}_{3,i1+1,i2}-E^{n}_{3,i1,i2}}{\Delta x_1}\\
\label{eq:yee2db3}B^{n+\frac{1}{2}}_{3,i1+\frac{1}{2},i2+\frac{1}{2}}=&~B^{n-\frac{1}{2}}_{3,i1+\frac{1}{2},i2+\frac{1}{2}}-c\Delta t\times \frac{E^{n}_{2,i1+1,i2+\frac{1}{2}}-E^{n}_{2,i1,i2+\frac{1}{2}}}{\Delta x_1}\nonumber\\
&+c\Delta t\times \frac{E^{n}_{1,i1+\frac{1}{2},i2+1}-E^{n}_{1,i1+\frac{1}{2},i2}}{\Delta x_2}
\end{align}
\begin{align}
\label{eq:yee2de1}E^{n+1}_{1,i1+\frac{1}{2},i2}=&~E^{n}_{1,i1+\frac{1}{2},i2}-4\pi\Delta t\times j^{n+\frac{1}{2}}_{1,i1+\frac{1}{2},i2}+c\Delta t\times \frac{B^{n+\frac{1}{2}}_{3,i1+\frac{1}{2},i2+\frac{1}{2}}-B^{n+\frac{1}{2}}_{3,i1+\frac{1}{2},i2-\frac{1}{2}}}{\Delta x_2}\\
\label{eq:yee2de2}E^{n+1}_{2,i1,i2+\frac{1}{2}}=&~E^{n}_{2,i1,i2+\frac{1}{2}}-4\pi\Delta t\times j^{n+\frac{1}{2}}_{2,i1,i2+\frac{1}{2}}-c\Delta t\times \frac{B^{n+\frac{1}{2}}_{3,i1+\frac{1}{2},i2+\frac{1}{2}}-B^{n+\frac{1}{2}}_{3,i1-\frac{1}{2},i2+\frac{1}{2}}}{\Delta x_1}\\
\label{eq:yee2de3}E^{n+1}_{3,i1,i2}=&~E^{n}_{3,i1,i2}-4\pi\Delta t\times j^{n+\frac{1}{2}}_{3,i1,i2}+c\Delta t\times \frac{B^{n+\frac{1}{2}}_{2,i1+\frac{1}{2},i2}-B^{n+\frac{1}{2}}_{2,i1-\frac{1}{2},i2}}{\Delta x_1}\nonumber\\
&~-c\Delta t\times \frac{B^{n+\frac{1}{2}}_{1,i1,i2+\frac{1}{2}}-B^{n+\frac{1}{2}}_{1,i1,i2-\frac{1}{2}}}{\Delta x_2}
\end{align}
where the EM field $\vec E$ and $\vec B$, and current $\vec j$ are defined with the proper half-grid offsets according to the Yee mesh \cite{Yee}. If we perform a Fourier transform of Eq. (\ref{eq:yee2db1})--(\ref{eq:yee2de3}) in both $x_1$ and $x_2$, and in time, Maxwell's equations reduce to 
\begin{align}
[\omega] \vec B= -[\vec k] \times \vec{E}\\
[\omega]\vec E= [\vec k] \times \vec{B}+4\pi \vec j
\end{align}
where
\begin{align}
[\vec k] = \biggl(\frac{\sin(k_1\Delta x_1/2)}{\Delta x_1/2},\frac{\sin(k_2\Delta x_2/2)}{\Delta x_2/2},0\biggr)\qquad [\omega]=\frac{\sin(\omega \Delta t/2)}{\Delta t/2}
\end{align}
In vacuum where $\vec j= 0$, the corresponding numerical dispersion relation for the EM waves is 
\begin{align}\label{eq:emdispersion}
[\omega]^2=c^2([k]^2_1+[k]^2_2)
\end{align}
The idea of a hybrid Yee-FFT solver is to keep the finite difference operator $[k]_2=\sin(k_2\Delta x_2/2)/(\Delta x_2/2)$ in the directions transverse to the drifting direction, while replacing the finite difference operator $[k]_1$ in the drifting direction with its spectral counterpart $[k]_1=k_1$. To achieve this, in the hybrid solver we will solve Maxwell's equations in $k_1$ space. The current is deposited locally using a rigorous charge conserving scheme that is equivalent to \cite {chargeconservation}. For the EM field and current, we first perform an FFT along $x_1$ so that all fields are defined in $(k_1,x_2)$ space. After that we apply a correction to the current in the drifting direction
\begin{align}\label{eq:currentcorrection}
\tilde j^{n+\frac{1}{2}}_1 = \frac{\sin{k_1\Delta x_1/2}}{k_1\Delta x_1/2}j^{n+\frac{1}{2}}_1
\end{align}
where $\tilde j_1$ is the corrected current. In \cite{VayJCP2013}, the current is also corrected where they combine a pure FFT solver with a charge conserving current deposit. This correction ensures that Gauss's Law is satisfied throughout the duration of the simulation if it is satisfied initially, as will be discussed in more detail in section \ref{sect:cc}. After the current correction we advance the EM field as
\begin{align}\label{eq:hybrid2db1}
B^{n+\frac{1}{2}}_{1,\kappa 1,i2+\frac{1}{2}}=&~B^{n-\frac{1}{2}}_{1,\kappa 1,i2+\frac{1}{2}}-c\Delta t\times \frac{E^{n}_{3,\kappa 1,i2+1}-E^{n}_{3,\kappa 1,i2}}{\Delta x_2}\\
\label{eq:hybrid2db2}B^{n+\frac{1}{2}}_{2,\kappa 1,i2}=&~B^{n-\frac{1}{2}}_{2,\kappa 1,i2}-i\xi^+k_1c\Delta t E^{n}_{3,\kappa 1,i2}\\
\label{eq:hybrid2db3}B^{n+\frac{1}{2}}_{3,\kappa 1,i2+\frac{1}{2}}=&~B^{n-\frac{1}{2}}_{3,\kappa 1,i2+\frac{1}{2}}+i\xi^+k_1c\Delta t E^{n}_{2,\kappa 1,i2+\frac{1}{2}}+c\Delta t\times \frac{E^{n}_{1,\kappa 1,i2+1}-E^{n}_{1,\kappa 1,i2}}{\Delta x_2}\\
\label{eq:hybrid2de1}E^{n+1}_{1,\kappa 1,i2}=&~E^{n}_{1,\kappa 1,i2}-4\pi\Delta t\times \tilde j^{n+\frac{1}{2}}_{1,\kappa 1,i2}+c\Delta t\times \frac{B^{n+\frac{1}{2}}_{3,\kappa 1,i2+\frac{1}{2}}-B^{n}_{3,\kappa 1,i2-\frac{1}{2}}}{\Delta x_2}\\
\label{eq:hybrid2de2}E^{n+1}_{2,\kappa 1,i2+\frac{1}{2}}=&~E^{n}_{2,\kappa 1,i2+\frac{1}{2}}-4\pi\Delta t\times j^{n+\frac{1}{2}}_{2,\kappa 1,i2+\frac{1}{2}}+i\xi^-k_1c\Delta t B^{n+\frac{1}{2}}_{3,\kappa 1,i2+\frac{1}{2}}\\
\label{eq:hybrid2de3}E^{n+1}_{3,\kappa 1,i2}=&~E^{n}_{3,\kappa 1,i2}-4\pi\Delta t\times j^{n+\frac{1}{2}}_{3,\kappa 1,i2}-i\xi^-k_1c\Delta t B^{n+\frac{1}{2}}_{2,\kappa 1,i2}\nonumber\\
&~-c\Delta t\times \frac{B^{n}_{1,\kappa 1,i2+\frac{1}{2}}-B^{n}_{1,\kappa 1,i2-\frac{1}{2}}}{\Delta x_2}
\end{align}
where $k_1=2\pi\kappa_1/N$ and $N$ is the number of grids in $x_1$ direction, and $\kappa_1=0,1,\ldots,N/2-1$ is the mode number. Note in the hybrid solver, the EM fields $\vec E$, $\vec B$, and current $\vec j$ have the same temporal and spatial centering as in the Yee solver, and
\begin{align}
\xi^\pm=\exp\biggl(\pm \frac{k_1\Delta x_1}{2}i\biggr )
\end{align}
is the phase shifting due to the half grid offsets of the $E_1$, $B_{2,3}$, and $j_1$ in the $\hat 1$-direction. Compared with the standard Yee solver algorithm, it is evident that if we replace $-ik_1$ with the corresponding finite difference form we can recover the standard 2D Yee algorithm. 

\subsection{Courant condition}
The Courant condition of the hybrid solver can be easily derived from the corresponding numerical EM dispersion Eq. (\ref{eq:emdispersion}). Substituting into Eq. (\ref{eq:emdispersion}) the finite difference operator in time $[\omega]$
\begin{align}
[\omega]=\frac{\sin(\omega\Delta t/2)}{\Delta t/2}
\end{align}
and the finite difference operators in space
\begin{align}
[k]_1=k_1\qquad [k]_2=\frac{\sin(k_2\Delta x_2/2)}{\Delta x_2/2}
\end{align}
we can obtain the corresponding constraint on the time step
\begin{align}
\frac{\Delta t}{2}\sqrt{k^2_1+\biggl(\frac{\sin(k_2\Delta x_2/2)}{\Delta x_2/2}\biggr)^2}\le 1
\end{align}
Note the $\vec k$ range of the fundamental Brillouin zone is $\vert k_1\vert \le \pi/\Delta x_1$, $\vert k_2\vert \le \pi/\Delta x_2$, we can obtain the Courant limit on the hybrid solver
\begin{align}
\Delta t \le \frac{2}{\sqrt{\frac{\pi^2}{\Delta x^2_1}+\frac{4}{\Delta x^2_2}}}
\end{align}
For square cells with $\Delta x_1=\Delta x_2$, this reduces to $\Delta t\le 0.537 \Delta x_1$.

\subsection{Charge conservation}
\label{sect:cc}
In the hybrid Yee-FFT solver, we rely on the Faraday's Law and Ampere's Law to advance the EM field. On the other hand, the local charge conserving current deposition \cite{chargeconservation} ensures the second-order-accurate finite difference representation of the continuity equation,
\begin{align}\label{eq:continuity}
\overline{\frac{\partial}{\partial t}}\rho^n_{i1,i2}+\frac{j^{n+\frac{1}{2}}_{1,i1+\frac{1}{2},i2}-j^{n+\frac{1}{2}}_{1,i1-\frac{1}{2},i2}}{\Delta x_1}+\frac{j^{n+\frac{1}{2}}_{2,i1,i2+\frac{1}{2}}-j^{n+\frac{1}{2}}_{2,i1,i2-\frac{1}{2}}}{\Delta x_2}=0
\end{align}
is satisfied, where
\begin{align}
\overline{\frac{\partial}{\partial t}}G^n=\frac{G^{n+1}-G^n}{\Delta t}
\end{align}
where $G^n$ is an arbitrary scalar quantity. Therefore, when combining this scheme with the second order accurate Yee solver, Gauss's Law is rigorously satisfied at every time step if it is satisfied at $t=0$. However, when the hybrid solver is used together with the charge conserving current deposition scheme, we need to apply a correction to the current, as shown in Eq. (\ref{eq:currentcorrection}), in order that the Gauss's Law is satisfied at every time step. This can be seen by first performing Fourier transform in the $x_1$ direction for Eq. (\ref{eq:continuity}),
\begin{align}\label{eq:continuityf}
\overline{\frac{\partial}{\partial t}}\rho^{n}_{\kappa 1,i2}-i\frac{\sin(k_1\Delta x_1/2)}{\Delta x_1/2}j^{n+\frac{1}{2}}_{\kappa 1,i2}+\frac{j^{n+\frac{1}{2}}_{2,\kappa 1,i2+\frac{1}{2}}-j^{n+\frac{1}{2}}_{2,\kappa 1,i2-\frac{1}{2}}}{\Delta x_2}=0
\end{align}
then applying the divergence operator of the hybrid solver to the left and right hand side of the Ampere's Law, Eq. (\ref{eq:hybrid2de1})--(\ref{eq:hybrid2de3}). Using Eq. (\ref{eq:continuityf}), we obtain 
\begin{align}
\overline{\frac{\partial }{\partial t}}\biggl(-4\pi \rho^n_{\kappa 1,i2}-ik_1E^{n}_{1,\kappa 1,i2}+\frac{E^{n}_{2,\kappa 1,i2+\frac{1}{2}}-E^{n}_{2,\kappa 1,i2-\frac{1}{2}}}{\Delta x_2}\biggr)=0
\end{align}
which shows that if  Gauss's Law for the 2D hybrid solver given by
\begin{align}
-ik_1E^{n}_{1,\kappa 1,i2}+\frac{E^{n}_{2,\kappa 1,i2+\frac{1}{2}}-E^{n}_{2,\kappa 1,i2-\frac{1}{2}}}{\Delta x_2}=4\pi \rho^n_{\kappa 1,i2}
\end{align}
is satisfied at $t=0$, it is satisfied at each time step. 

\subsection{3D Cartesian geometry}
It is straightforward to extend the hybrid solver to 3D cartesian geometry. In 3D Cartesian coordinates,  we solve Maxwell's equation in $(k_1, x_2, x_3)$ space where we use the same second order accurate finite difference form of the Yee solver in the $\hat 2$ and $\hat 3$ directions. As in the 2D Cartesian case, the current correction is applied to $j_1$ to ensure the Gauss's Law is satisfied. We have implemented the hybrid solver in 2D and 3D with current correction in our finite-difference-time-domain (FDTD) code OSIRIS \cite{osiris}.

\section{Numerical Cerenkov instability}
\label{sect:nci}
\begin{figure}[th]
\begin{center}\includegraphics[width=1\textwidth]{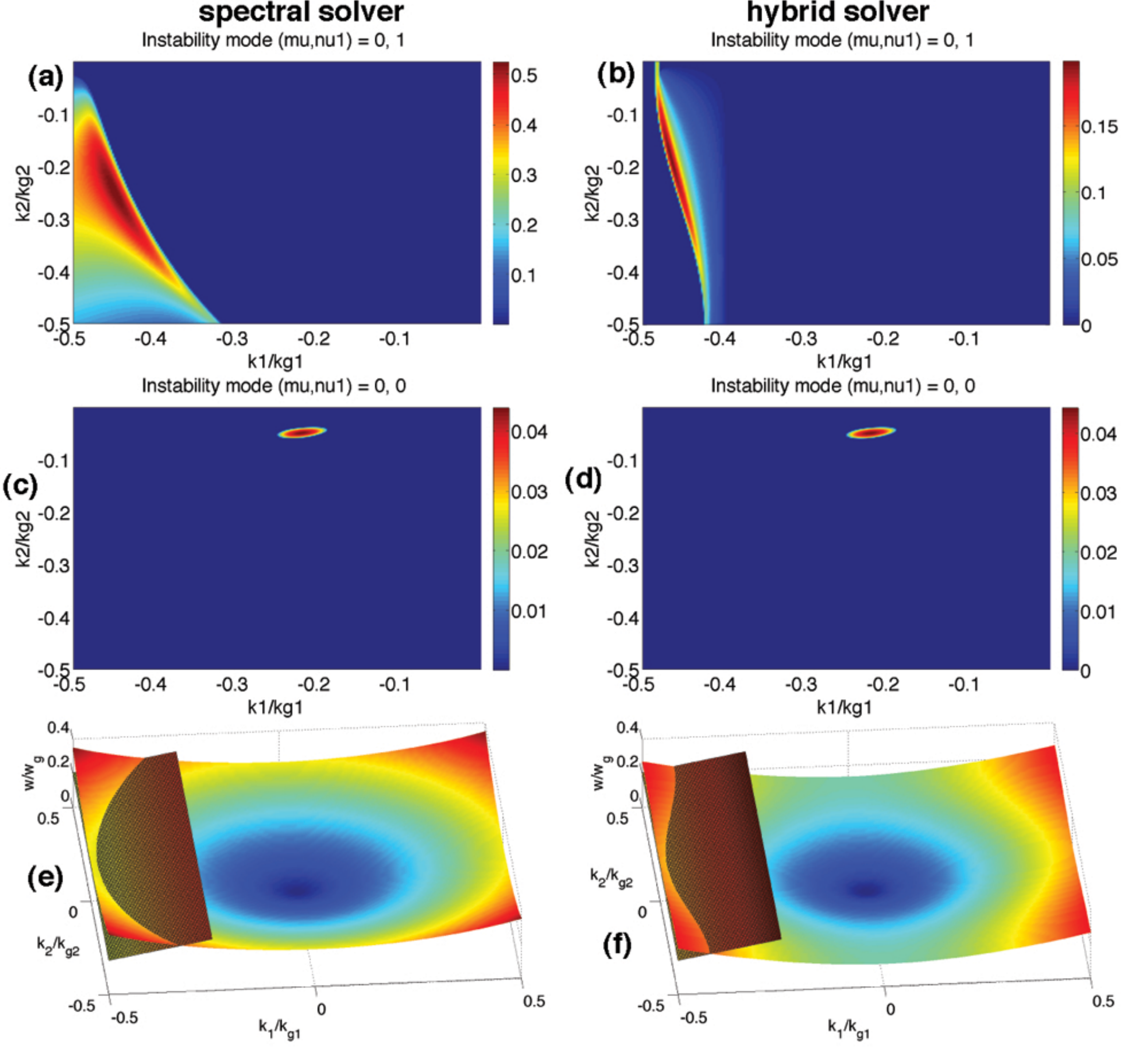}
\caption{\label{fig:embeam} The pattern of the $(\mu,\nu_1)=(0,\pm 1)$ modes for the two solvers are shown in (a) and (b). The pattern of the $(\mu,\nu_1)=(0,0)$ modes for  two solvers are shown in (c) and (d).  The intersection between the EM dispersion relations with the first spatial aliasing beam modes for the full spectral solver and the hybrid solver are shown in (e) and (f). When generating these plots we use $\Delta x_1=\Delta x_2=0.2~k^{-1}_0$, and $\Delta t=0.08~\omega^{-1}_0$. Other parameters are listed in Table \ref{tab:ncipara}. } 
\end{center}
\end{figure}

To investigate the NCI properties of the hybrid solver, we first consider its corresponding numerical dispersion relation. Employing the general theoretical framework established in Ref. \cite{XuCPC13,YuarXiv14}, we can calculate in detail the NCI modes for any Maxwell solver. The roots of the numerical dispersion relation that lead to the NCI can be found numerically by solving Eq. (17) in \cite{XuCPC13}, or by the analytical expression in Eq. (19) of \cite{YuarXiv14}. For convenience 
we present the corresponding numerical dispersion and analytical expressions of Eq. (17) of \cite{XuCPC13} in \ref{sect:appendix}. For the Yee solver the $k$ space representation of the finite difference operator is
\begin{align}
[k]_{i} = \frac{\sin(k_{i}\Delta x_{i}/2)}{\Delta x_{i}/2}
\end{align}
where $i=1,2$ in 2D. Meanwhile, in the hybrid solver the $\vec k$ space operator in the drifting direction is replaced with that of the spectral solver $[k]_1 \rightarrow k_1$. By substituting the respective operators for each direction into Eq. (19) of Ref. \cite{YuarXiv14} [or Eq. (\ref{eq:asym1}) in \ref{sect:appendix}], we can rapidly find the set of NCI modes for the hybrid solver. In Fig. \ref{fig:embeam} (a)--(d), we plot the $(\mu,\nu_1)=(0,0)$  and $(\mu,\nu_1)=(0, \pm 1)$ modes for the hybrid  and spectral solvers by scanning over the $(k_1,k_2)$ space in the fundamental Brillouin zone and solve for the growth rates of the corresponding unstable modes. The parameters used to generate this plot are listed in Table \ref{tab:ncipara}. 

\begin{table}[t]
\centering
\begin{tabular}{p{8cm}c}
\hline\hline
\textbf{Parameters} & \textbf{Values}\\ 
\hline
grid size $(k_p\Delta x_1, k_p\Delta x_2)$ & $(0.2,0.2)$\\
time step $\omega_p\Delta t$ & $0.4\Delta x_1$\\
boundary condition & Periodic \\
simulation box size $(k_pL_1,k_pL_2)$ & 51.2$\times 51.2$\\
plasma drifting Lorentz factor & $\gamma=50.0$\\
plasma density & $n/n_p = 100.0$\\
\hline\hline
\end{tabular}
\caption{Crucial simulation parameters for the 2D relativistic plasma drift simulation. $n_p$ is the reference plasma density, and $\omega^2_p=4\pi q^2n_p/m_e$, $k_p=\omega_p$ ($c$ is normalized to 1). }
\label{tab:ncipara}
\end{table}

\begin{figure}[th]
\begin{center}\includegraphics[width=1\textwidth]{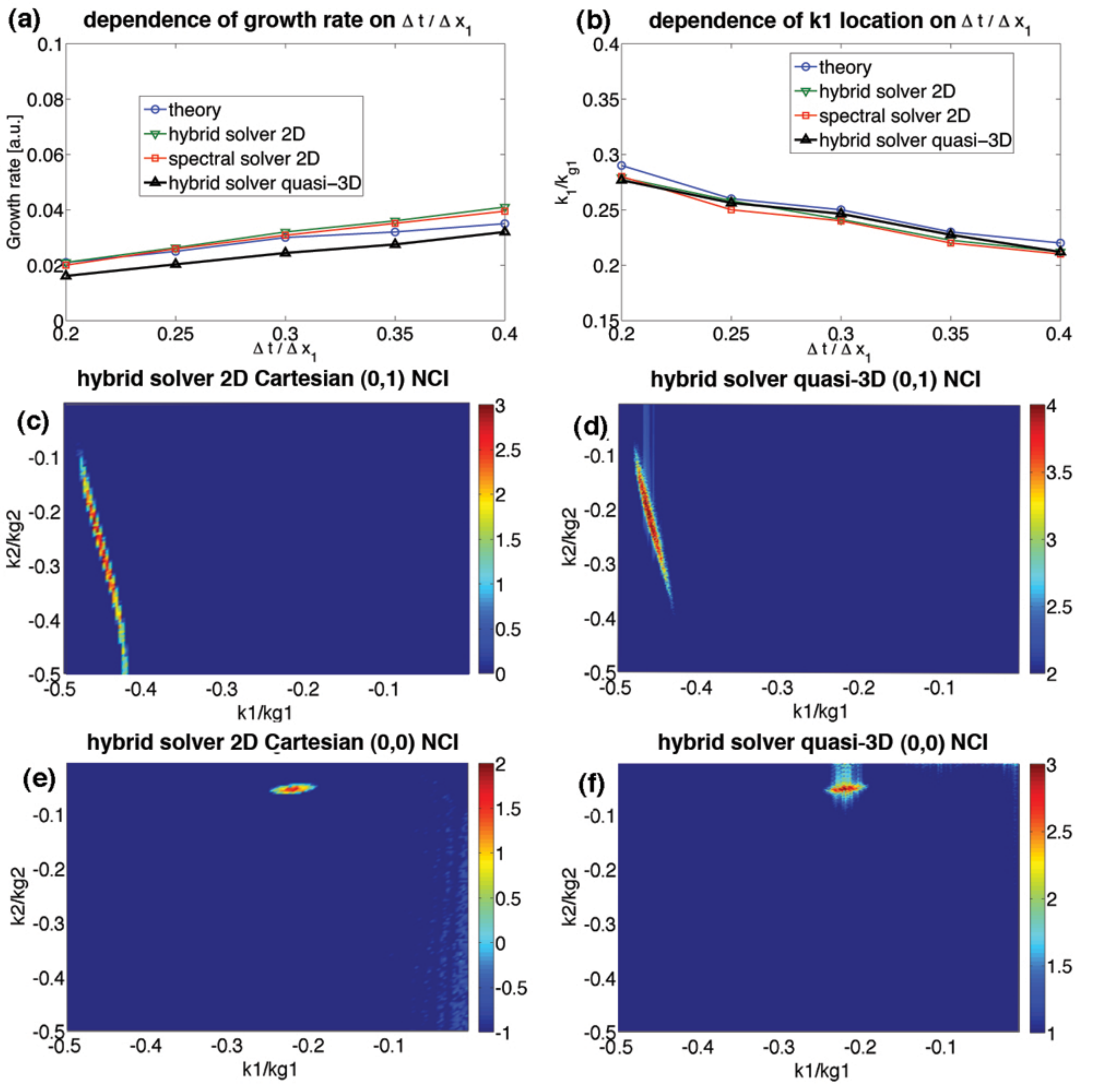}
\caption{\label{fig:ncigr} In (a) and (b) the dependence of the growth rate and  $k_1$ for the fastest growing $(\mu,\nu_1)=(0,0)$ mode on the time step is shown. The four lines correspond to the theoretical prediction for the hybrid solver in 2D,  results from OSIRIS and UPIC-EMMA simulations for the spectral and hybrid solvers in 2D Cartesian geometry, and results for the hybrid solver in the quasi-3D geometry (where the $k_2$ is obtained from a Hankel transform). In (c)--(f)  the spectrum of $E_2$ ($E_\rho$) is plotted for OSIRIS simulations with the hybrid solver in 2D Cartesian or the quasi-3D geometry. In (c) and (d) results from runs where no filter in $k_1$ is used to eliminate the $(\mu,\nu_1)=(0,\pm 1)$ modes. In (e) and (f) a filter in $k_1$ was used to eliminate the $(\mu,\nu_1)=(0,\pm 1)$ modes and now the $(\mu,\nu_1)=(0,0)$ modes are seen. These results show that the 2D Cartesian and quasi-3D geometries have very similar properties and that the strategies used to eliminate the NCI in 2D Cartesian can be applied to the quasi-3D case.  }
\end{center}
\end{figure}

We can see from Fig. \ref{fig:embeam} (a) and (b) that the $(\mu,\nu_1)=(0,\pm 1)$ NCI modes of the two solvers reside near the edge of the fundamental Brillouin zone, although the patterns are slightly different due to their different finite difference operators in the $\hat 2$-direction, which leads to the slightly different EM dispersion curves. In Fig. \ref{fig:embeam} (e) and (f) we show how different EM dispersion curves leads to different $(\mu,\nu_1)=(0,\pm 1)$ NCI modes for the two solvers. These modes are distinct, and far removed from the modes of physical interest, and are relatively easy to eliminate.

More importantly, we see from Fig. \ref{fig:embeam} (c) and (d) that the hybrid solver leads to $(\mu,\nu_1)=(0,0)$ NCI modes that are very similar to their spectral solver counterpart. The pattern of the $(\mu,\nu_1)=(0,0)$ modes for these two solvers are both four dots (in 2D) and highly localized in the fundamental Brillouin zone. We also use the theory to perform parameter scan to study the dependence of growth rates (of the fastest growing mode) and the locations in $k$ space of the $(\mu,\nu_1)=(0,0)$ modes on $\Delta t/\Delta x_1$ for the hybrid solver, and compare this result against that of the fully spectral solver, as shown in Fig. \ref{fig:ncigr} (a) and (b). We likewise carried out OSIRIS simulations using the hybrid solver and UPIC-EMMA \cite{YuJCP14,upic} using the spectral solver, to compare against theoretical results. Very good agreement is found between theory and simulations. Fig. \ref{fig:ncigr} (a) and (b) show that both the $k_1$ location, and growth rates of the $(\mu,\nu_1)=(0,0)$ modes are almost identical for the two solvers. This indicates that, just like the spectral solver, the growth rate of the $(\mu,\nu_1)=(0,0)$ modes of the hybrid solver is reduced, while their location in $k_1$ increases when the time step is reduced. 

In Fig. \ref{fig:ncigr} (c) and (e) we show the locations of the unstable $(\mu,\nu_1)=(0,\pm 1)$, and $(\mu,\nu_1)=(0,0)$ NCI modes for the hybrid solver in OSIRIS for 2D Cartesian geometry. The agreement between Fig. \ref{fig:ncigr} (c) and Fig. \ref{fig:embeam} (b), and between Fig. \ref{fig:ncigr} (e) and Fig. \ref{fig:embeam} (d) are excellent.

The main advantage of the purely spectral solver regarding its NCI properties in comparison to a purely FDTD solver is that the superluminal dispersion relation makes it much easier to eliminate the NCI modes at $(\mu,\nu_1)=(0,0)$: the modes have a growth rate that is one order of magnitude smaller than that for the $(\mu,\nu_1)=(0,\pm 1)$ modes, their locations are highly localized in $\vec k$ space, and they can be moved away from the modes of physical interest by reducing the time step. We showed above that similar NCI properties can be achieved by using a hybrid FDTD-spectral solver, where the Maxwell's equation are solved in Fourier space only in the direction of the plasma drift. Comparing with an EM-PIC code using a multi-dimensional spectral solver which solves Maxwell's equation in $\vec{k}$ space, there are advantages when solving it in $(k_1,x_2)$ space in 2D [and $(k_1,x_2,x_3)$ space in 3D]. Firstly, the hybrid solver saves the FFT in the other directions; secondly, since the solver is FDTD in the directions transverse to the drifting direction, it is easier to integrate the algorithm into existing FDTD codes such as OSIRIS where the parallelizations and boundary conditions in the transverse direction can remain untouched. Last but perhaps most important, the idea that one can obtain preferable NCI properties by solving Maxwell's equation in $k_1$ space in the drifting direction can be readily extended to the quasi-3D algorithm \cite{quasi3d}, as we can solve the Maxwell's equation in $(k_1, \rho,\psi)$ space. 

\section{Elimination of the NCI modes}
\label{sect:nci00clean}
In Ref. \cite{YuarXiv14}, we proposed strategies to eliminate the NCI in the spectral solver. These strategies can be readily applied to the hybrid solver. For square (or cubic) cell, the pattern of the fastest growing modes resides in a narrow range of $k_1$ near the edge of the fundamental Brillouin zone. Therefore we can apply a low-pass filter in $k_1$ to the current to eliminate the fastest growing modes. Since the fields are already in $k_1$ space when solving the Maxwell's equations, the filtering can be done efficiently by applying a form factor to the current only in $k_1$. 

\begin{figure}[th]
\begin{center}\includegraphics[width=1\textwidth]{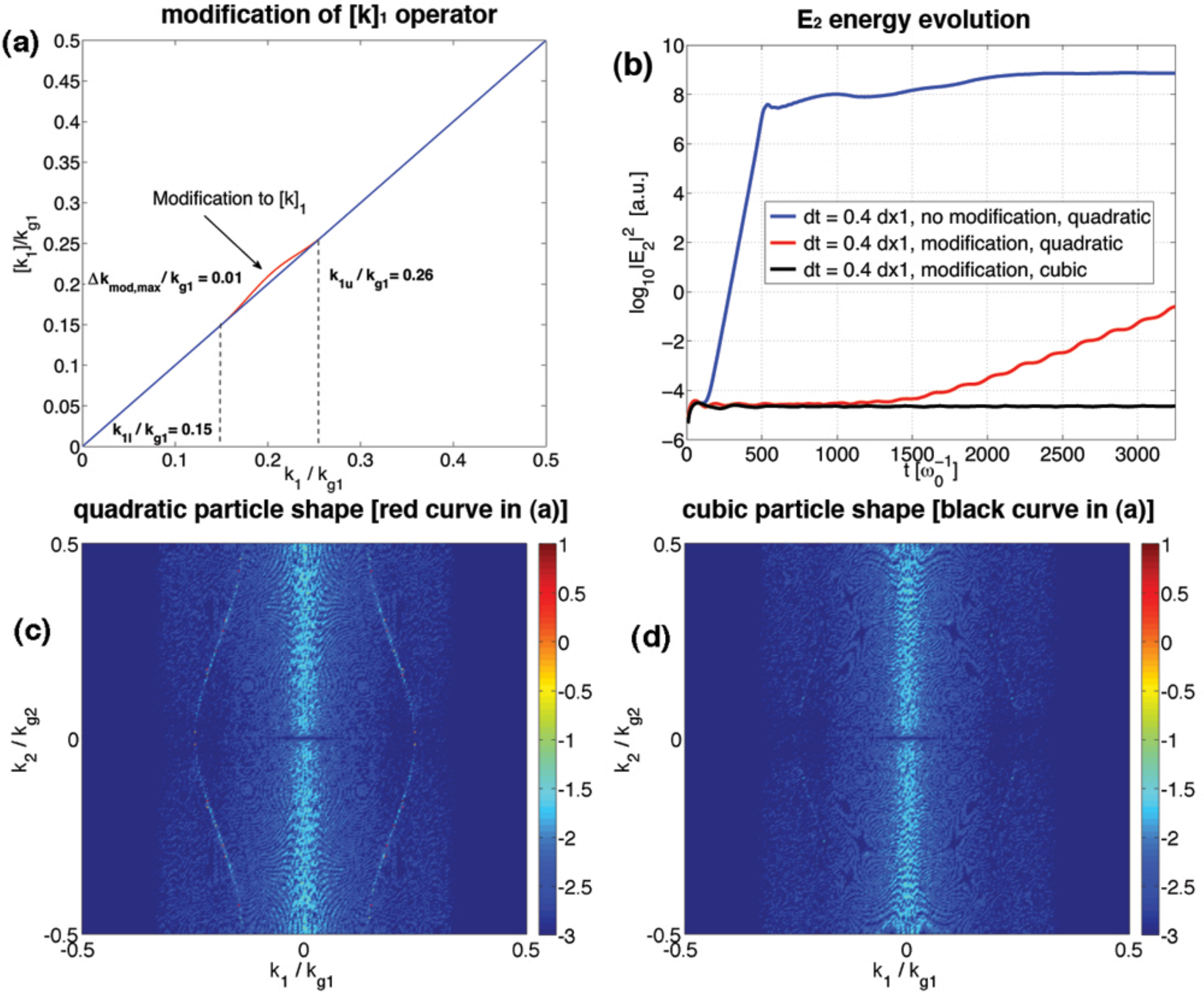}
\caption{\label{fig:nci00clean} In (a) the perturbation to $[k]_1$ that is used to eliminate the $(\mu,\nu_1)=(0,0)$ NCI modes is shown. In (b) the evolution of the  $\log_{10}\vert E_2\vert^2$  for a reference case and for two cases with the EM dispersion modification (one with quadratic and another with cubic particle shapes). In (c) and (d), the  spectrum of $E_2$ at $t=3200~\omega^{-1}_0$ is shown for the two cases with the EM dispersion modifications. In (c) quadratic particle shapes are used, while in (d) cubic particle shapes are used.} 
\end{center}
\end{figure}

As for the $(\mu,\nu_1)=(0,0)$ mode, if they are near the main or higher order harmonics of the physical modes, we can move them away and reduce their growth rates by simply reducing the time step. To further mitigate the $(\mu,\nu_1)=(0,0)$ NCI modes when they are far away from the physical modes, one can modify the EM dispersion relation, according to the procedure described in Ref. \cite{YuarXiv14}, to completely eliminate them. In Fig. \ref{fig:nci00clean} we plot how the modification is accomplished in the hybrid solver. As shown in Fig. \ref{fig:nci00clean} (a) except for the bump region for most $k_1$ the $[k]_1$ for a particular $k_1$ is $k_1$ itself; near the bump, the $[k]_1$ for $k_1$ is $k_1+\Delta k_{mod}$, where $\Delta k_{mod}$ is a function of $k_1$ with
\begin{align}
\Delta k_{mod}= \Delta k_{mod,\max}\cos\biggl(\frac{k_{1}-k_{1m}}{k_{1l}-k_{1m}}\frac{\pi}{2}\biggr)^2
\end{align}
 where $k_{1l}$, $k_{1u}$ are the lower and upper $k_1$ to be modified, $k_{1m}=(k_{1l}+k_{1u})/2$, and $\Delta k_{mod,\max}$ is the maximum value of $\Delta k_{mod}$. The values of $k_{1l}$, $k_{1u}$ and $\Delta k_{mod,\max}$ are determined by the position of the $(\mu,\nu_1)=(0,0)$ modes and their growth rates. According to the NCI theory, for the parameters in Table \ref{tab:ncipara}, when the $[k]_1$ is as defined in Fig. \ref{fig:nci00clean} (a) (with $k_{1l}/k_{g1}=0.15$, $k_{1u}/k_{g1}=0.26$, and $\Delta k_{mod,\max}/k_{g1}=0.01$), there is no unstable $(\mu,\nu_1)=(0,0)$ NCI modes, i.e., the $(\mu,\nu_1)=(0,0)$ mode has a theoretical growth rate of zero. To verify the theoretical results in the hybrid solver, in Fig. \ref{fig:nci00clean} (b) we plot the $E_2$ energy growth with and without the modification.
 In these simulations we used the parameters in Table \ref{tab:ncipara}. The blue curve in Fig. \ref{fig:nci00clean} (b) represents the case without the modification, while the red and black curves are those with the modification to $k_1$. The cases with blue and red curves used quadratic particle shapes, while the case for the black curve used cubic particle shapes. We have likewise plotted the $E_2$ spectra at the time point $t=3200~\omega^{-1}_0$ indicated in Fig. \ref{fig:nci00clean} (c) and (d) for the two cases with the modifications (red and black curves in \ref{fig:nci00clean} (b)).  We can see from Fig. \ref{fig:nci00clean} (b) and (c) that after the modification, the growth rate of the $(\mu,\nu_1)=(0,0)$ NCI modes reduces to zero. Meanwhile, the  red curve rises later in time due to the $(\mu,\nu_1)=(\pm 1,\pm 2)$ NCI modes. As we showed in Ref. \cite{YuarXiv14} the growth rate of these higher order modes can be reduced by using higher order particle shape. Therefore when cubic particle shapes are used, as is the case for the black curve, the $(\mu,\nu_1)=(\pm 1,\pm 2)$ NCI modes do not grow exponentially and are therefore much less observable in the corresponding spectrum at $t=3200~\omega^{-1}_0$ in Fig. \ref{fig:nci00clean} (d) as compared to \ref{fig:nci00clean} (c).

\section{hybrid solver in quasi-3D algorithm}
\label{sect:quasi3dhybrid}
As mentioned in section \ref{sect:intro}, the idea of the hybrid solver can be easily incorporated into the quasi-3D algorithm \cite{quasi3d,davidson} in which the fields and current are expanded into azimuthal Fourier modes. 
We can obtain the hybrid Yee-FFT solver for the quasi-3D algorithm by using FFTs in the z ($x_1$) direction and finite difference operators in r ($x_2$) direction in the equations for each azimuthal mode. Note in quasi-3D OSIRIS we use a charge conserving current deposition scheme for the Yee solver (as described in \cite{davidson}), therefore for the hybrid solver adapted for the quasi-3D algorithm we can apply the same current correction for the use of FFTs to $j_1$ in order that the Gauss's Law is satisfied throughout the duration of the simulation. 

The NCI properties of the hybrid solver for the quasi-3D algorithm are similar to that of the 2D Cartesian geometry \cite{YuAAC14}. While a rigorous NCI theory for the quasi-3D algorithm is still under development, we can empirically investigate the NCI for this geometry through simulation. In Fig. \ref{fig:ncigr} (d) and (f) we plot the $E_r$ data at a time during the exponential growth of the EM fields due to the NCI, which shows the $(\mu,\nu_1)=(0,\pm 1)$ and $(\mu,\nu_1)=(0,0)$ modes for the hybrid solver in quasi-3D geometry. For the $E_r$ data, we conduct an FFT  in $x_1$ and a Hankel transform in $x_2$. Similarly to the 2D Cartesian case, we isolate the $(\mu,\nu_1)=(0,0)$ modes by applying a low-pass filter in the current in $k_1$ space to eliminate the fastest growing $(\mu,\nu_1)=(0,\pm 1)$ NCI modes. The parameters used in the simulations are listed in Table \ref{tab:ncipara}, and a conducting boundary is used for the upper $r$ boundary. We kept azimuthal modes of $m=-1,0,1$ in the simulations.

By comparing Figs. \ref{fig:ncigr} (c)--(f) it can be seen that the pattern of the NCI modes are similar for the ($x_2, x_1$) and ($r,z$) geometries.  We have also plotted the dependence of the growth rate and $k_1$ position of the $(\mu,\nu_1)=(0,0)$ NCI modes for the quasi-3D geometry in Fig.  \ref{fig:ncigr} (a) and (b). These plots show that when the time step decreases the growth rates of the $(\mu,\nu_1)=(0,0)$ NCI modes in the quasi-3D geometry decreases, while the  $k_1$ position increases (and move away from the physical modes), in a nearly similar fashion to 2D Cartesian geometry. This indicates that the same strategies for eliminating NCI in 2D Cartesian geometry can be applied to the quasi-3D geometry. The fastest growing modes residing at the edge of the fundamental Brillouin zone can be eliminated by applying a low-pass filter in the current. The $(\mu,\nu_1)=(0,0)$ NCI modes can be mitigated by either reducing the time step to lower the growth rate and move the modes away from the physics in $k_1$ space, or by modifying the $[k]_1$ operator as discussed in section \ref{sect:nci00clean} to create a bump in the EM dispersion along the $k_1$ direction. We have implemented the modification to the $[k]_1$ operator  into the hybrid solver for the quasi-3D OSIRIS code, and have confirmed that this modification completely eliminate the $(\mu,\nu_1)=(0,0)$ NCI modes.  The coefficients used for the modification are the same as those for the 2D Cartesian case discussed in section \ref{sect:nci00clean}.  

\section{Sample simulations}
\label{sect:lwfa}
\begin{figure}[th]
\begin{center}\includegraphics[width=1\textwidth]{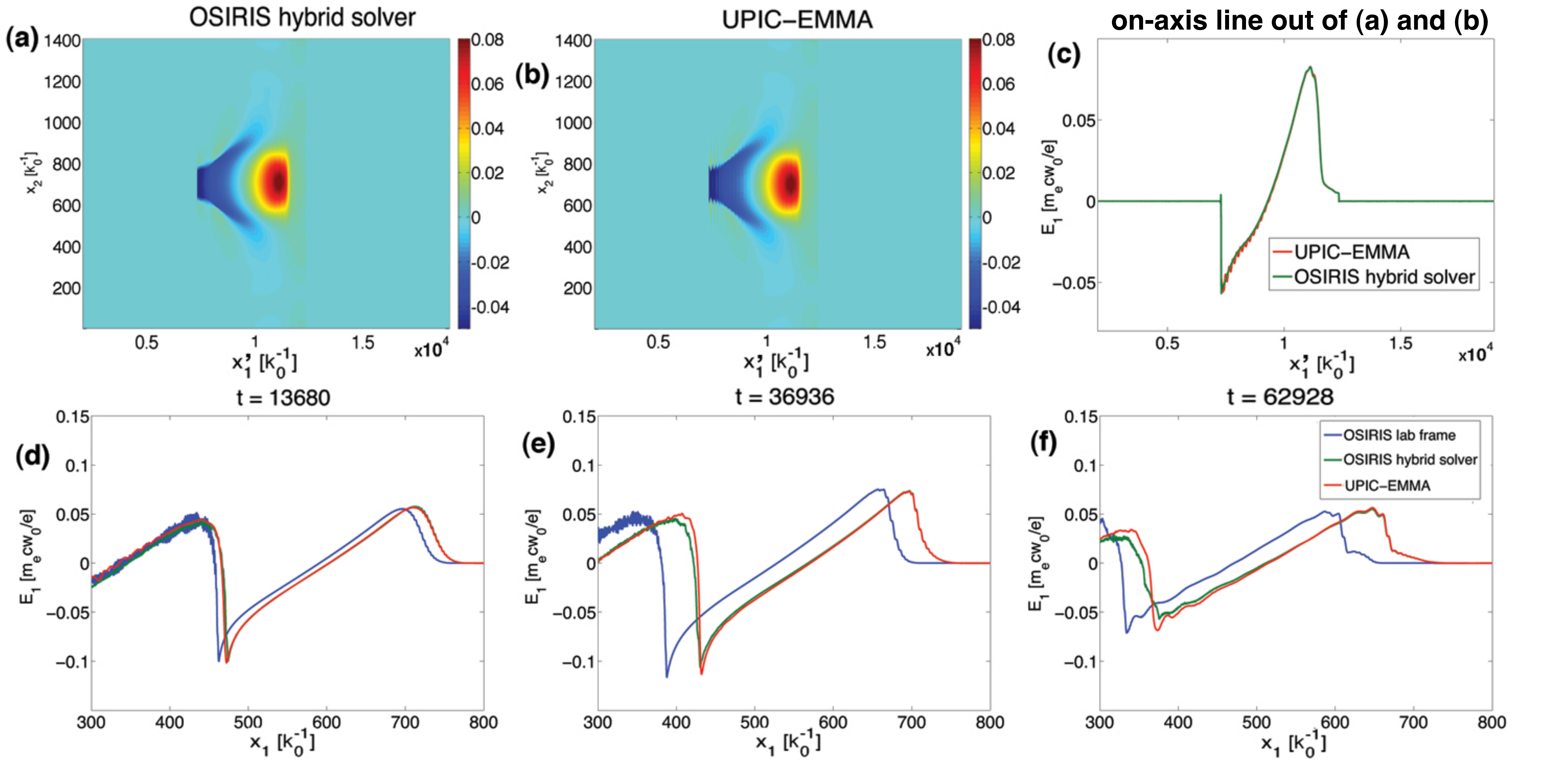}
\caption{\label{fig:lwfae1} Comparison between OSIRIS lab frame, OSIRIS with the hybrid solver in the boosted frame and UPIC-EMMA in the boosted frame. In (a) and (b),  2D plots of $E_1$  for OSIRIS with the hybrid solver and UPIC-EMMA at $t'=3955\omega^{-1}_0$ are shown in the boosted frame, where $\omega_0$ is the laser frequency in the lab frame. In (c), lineouts along the laser propagation direction of the same data are shown. In (d)--(f), lineouts of the $E_1$ data transformed back to the lab frame are shown. The colored lines correspond to an OSIRIS lab frame simulation, an OSIRIS hybrid solver simulation in the Lorentz boosted frame, and UPIC-EMMA simulation in the Lorentz boosted frame.} 
\end{center}
\end{figure}

In this section, we present preliminary results of Lorentz boosted frame LWFA simulations using the hybrid solver in OSIRIS. For comparison, we performed simulations with the same parameters using UPIC-EMMA which uses a spectral Maxwell solver. Table \ref{tab:lwfapara1} lists the simulation parameters. We use a moving antenna in both cases to launch lasers into the plasma. The results are summarized in Fig. \ref{fig:lwfae1}. 

In Fig. \ref{fig:lwfae1} (a)--(b) the $E_1$ field at $t'=3955\omega^{-1}_0$ for simulations with both the hybrid solver and spectral solver in the Lorentz boosted frame are plotted, where $\omega_0$ is the laser frequency in the lab frame. Both the spectral solver and hybrid solver give similar boosted frame results, and there is no evidence of NCI affecting the physics in either case. We plot the line out of the on-axis wakefield in Fig. \ref{fig:lwfae1} (c), which shows very good agreement with one another. The very good agreement can also be seen when we transformed the boosted frame data back to the lab frame. In Fig. \ref{fig:lwfae1} (d)--(f) we plot the on-axis $E_1$ field for the OSIRIS lab frame data, the transformed data for the OSIRIS boosted frame simulation with the hybrid solver, and the transformed data from theUPIC-EMMA boosted simulation  at several  values of time  in the lab frame. As seen in Fig. \ref{fig:lwfae1} (d)--(f), the transformed data from the two boosted frame simulations agrees very well with each other. Note the displacement of the lineouts between the lab frame data and boosted frame data is due to the different group velocity of the laser between the Yee, spectral, and hybrid solver. Since the finite difference operator $[k]_1$ is the same for the spectral and hybrid solver, the group velocity of the laser along its propagation direction is the same for the two boosted frame simulations. As a result the on-axis transformed laser data of the two boosted frame simulations almost reside on top of one another and are more accurate. 

\begin{table}[t]
\centering
\begin{tabular}{lr}
\hline\hline
Plasma &\\
\quad density $n_0$& $1.148\times 10^{-3} n_0\gamma_b$\\
\quad length $L$ & $7.07\times 10^4k^{-1}_0/\gamma_b$\\
Laser & \\
\quad pulse length $\tau$ & $ 70.64k^{-1}_0\gamma_b(1+\beta_b)$\\
\quad pulse waist $W$ & $117.81k^{-1}_0$\\
\quad polarization & $\hat 3$-direction\\
\quad normalized vector potential $a_0$ & 4.0\\
2D boosted frame simulation&\\
\quad grid size $\Delta x_{1,2}$ & $0.0982k^{-1}_0\gamma_b(1+\beta_b)$\\
\quad time step $\Delta t/\Delta x_1$ &0.225\\
\quad number of grid $(\gamma_b=14)$& 8192$\times$512\\
\quad particle shape & quadratic\\
\hline\hline
\end{tabular}
\caption{Parameters for a 2D LWFA  simulations in a Lorentz boosted frame that were used for in 2D Cartesian geometry with the  hybrid solver in OSIRIS and with a fully spectral solver in UPIC-EMMA. The laser frequency $\omega_0$ and  number $k_0$ in the lab frame are used to normalize simulation parameters. The density is normalized to the critical density in the lab frame, $n_0=m_e\omega^2_0/(4\pi e^2)$.}
\label{tab:lwfapara1}
\end{table}

\section{Summary}
\label{sect:summary}
We proposed to use a hybrid Yee-FFT and a rigorous charge conserving current deposit for solving Maxwell's equations in order to eliminate the numerical Cerenkov instability in PIC codes when modeling plasmas or beams that drift with relativistic speeds in a particular direction. In this solver we solve the Maxwell's equation in $k_1$ space along the drifting direction ($\hat x_1$ direction), and use second order finite difference representation for the derivatives in the other directions. This provides greater than $N$-th order accuracy for the spatial derivatives in the $\hat x_1$ direction, while keeping the locality of the field solve and current deposit in the directions transverse to $\hat 1$. For the current deposit, we start from the charge conserving deposit in  OSIRIS and then correct it so that it still satisfies the continuity equation for the hybrid solver. Thus, Gauss's law remains rigorously satisfied at every time step if it is satisfied initially. 

It is found from the NCI theory that such a hybrid solver has similar NCI properties in comparison to a full spectral solver that solves Maxwell equation in multi-dimensional $\vec k$ space. As a result, the $(\mu,\nu_1)=(0,0)$ NCI modes have a growth rate one order of magnitude smaller than the fastest growing $(\mu,\nu_1)=(0,\pm 1)$ NCI modes, and are highly localized. In addition, the growth rates of the $(\mu,\nu_1)=(0,0)$ modes decrease as one reduces the simulation time step, and their locations in Fourier space also move farther away from the physics. 

Compared with the spectral solver, the hybrid solver performs an FFT only along the drifting direction of the plasma. As a result, it saves the computation of FFT in the other directions if this ultimately becomes an issue for parallel scalability. In addition, it can be readily adapted into fully operational FDTD codes without the need to modify various boundary conditions in the transverse directions. Very importantly, this idea can be readily applied to the quasi-3D algorithm in which the quantities are decomposed into azimuthal harmonics. In this algorithm FFTs cannot be used in the $\hat r$ direction. We demonstrate the feasibility of the hybrid Yee-FFT solver in 2D/3D Cartesian geometry, as well as in the quasi-3D geometry. Although we have not conducted a rigorous theoretical analysis for the NCI in the r-z or quasi-3D geometries, we find in simulations that the hybrid solver in quasi-3D geometry has very similar NCI properties to that in the 2D Cartesian geometry. 

We show that the strategy to eliminate NCI in the hybrid solver for 2D/3D Cartesian geometry, as well as quasi-3D geometry, is similar to that for the spectral solver. The fastest growing NCI modes can be eliminated by applying a low-pass filter in the current. The $(\mu,\nu_1)=(0,0)$ NCI modes can be eliminated by reducing the time step which both reduces their growth rates and moves them away from the physical modes in Fourier space. These NCI modes can also be fully eliminated by slightly modifying the the EM dispersion relation along $k_1$ direction at the location in Fourier space where the $(\mu,\nu_1)=(0,0)$ modes reside. This approach is demonstrated in both Cartesian and quasi-3D geometry. 

We showed that  the new hybrid solver in OSIRIS can be used to conduct 2D LWFA simulations in a Lorentz boosted frame. With the low-pass filter applied to current and using reduced time step, we observe no evidence of NCI affecting the physics in the simulation. Very good agreement is found between the results from OSIRIS with the hybrid solver,   UPIC-EMMA simulations,  as well as OSIRIS lab frame simulations with the standard Yee solver. This demonstrates the feasibility of using the hybrid solver to perform high fidelity relativistic plasma drift simulation. 

This work was supported by US DOE under grants DE-SC0008491, DE-SC0008316, DE-FC02-04ER54789, DE-FG02-92ER40727, by the US National Science Foundation under the grant ACI 1339893, and by NSFC Grant 11175102, thousand young talents program, and by the European Research Council (ERC-2010-AdG Grant 267841), and by LLNL's Lawrence Fellowship. Simulations were carried out on the UCLA Hoffman2 and Dawson2 Clusters, and on Hopper cluster of the National Energy Research Scientific Computing Center.

\begin{appendix}
\section{Numerical dispersion for relativistically drifting plasma and NCI analytical expression in hybrid solver}
\label{sect:appendix}
According to Ref. \cite{XuCPC13,YuarXiv14}, the numerical dispersion for the hybrid solver can be expressed as
\begin{align}\label{eq:2dmodesres}
& \left((\omega' - k'_1 v_0)^2- \frac{\omega_p^2}{\gamma^3}(-1)^\mu\frac{S_{j1} S_{E1}\omega'}{[\omega]} \right)\times  \nonumber\\
&\left( [\omega]^2 - [k]_{E1}[k]_{B1}  - [k]_{E2}[k]_{B2}  - \frac{\omega_p^2}{\gamma} (-1)^\mu\frac{S_{j2}(S_{E2}[\omega] - S_{B3} [k]_{E1} v_0)}{\omega' - k'_1 v_0} \right) \nonumber\\
&+ \mathcal{C}=0
\end{align}
where $\mathcal{C}$ is a coupling term in the dispersion relation
\begin{align}\label{eq:coupling}
\mathcal{C}=\frac{\omega_p^2}{\gamma} \frac{(-1)^\mu}{[\omega]}\biggl\{&S_{j1}S_{E1}\omega'[k]_{E2}[k]_{B2}(v^2_0-1)+S_{j2}S_{E2}[k]_{E2}[k]_{B2}(\omega'-k'_1v_0)\nonumber\\
&+S_{j1}[k]_{E2}(S_{E2}[k]_{B1}k_2v_0-S_{B3}k_2v^2_0[\omega])\biggr\}
\end{align}
and for the hybrid solver
\begin{align}
[k]_{E1}=[k]_{B1}=k_1\qquad [k]_{E2}=[k]_{B2}=\frac{\sin(k_2\Delta x_2/2)}{\Delta x_2/2}
\end{align}

We can expand $\omega'$ around the beam resonance $\omega' = k'_1v_0$ in Eq. (\ref{eq:2dmodesres}), and write $\omega' = k'_1v_0 + \delta\omega'$, where $\delta\omega'$ is a small term. This leads to a cubic equation for $\delta\omega'$ (see \cite{YuarXiv14} for the detailed derivation),
\begin{align}\label{eq:asym1}
A_2\delta\omega'^3+B_2\delta\omega'^2+C_2\delta\omega'+D_2=0
\end{align}
where
\begin{align}
A_2=&2\xi^3_0\xi_1\nonumber\\
B_2=&\xi^2_0\biggl\{\xi^2_0-[k]_{E1}[k]_{B1}-[k]_{E2}[k]_{B2}-\frac{\omega^2_p}{\gamma}(-1)^\mu S_{j2}(S_{E2}\xi_1-\zeta_1S'_{B3}[k]_{E1})\biggr\}\nonumber\\
C_2=&\frac{\omega^2_p}{\gamma}(-1)^\mu \biggl\{ \xi^2_0S_{j2}(\zeta_0S'_{B3}[k]_{E1}-S_{E2}\xi_0)-{\xi_1}S_{j1}[k]_{E2}k_2S_{E2}[k]_{B1}\nonumber\\
& +\xi_0[k]_{E2}(S_{j2}S_{E2}[k]_{B2}-S_{j1}k_2\zeta_1S'_{B3}\xi_0)\biggr\}\nonumber\\
\label{eq:asym2gen}D_2=&\frac{\omega^2_p}{\gamma}(-1)^\mu \xi_0[k]_{E2}k_2S_{j1}\biggl( S_{E2}[k]_{B1}-\zeta_0S'_{B3}\xi_0\biggr)
\end{align}
where
\begin{align}\label{eq:approximation}
\xi_0&=\frac{\sin(\tilde k_1\Delta t/2)}{\Delta t/2}\qquad \xi_1 = \cos(\tilde k_1\Delta t/2)\nonumber\\
\zeta_0&= \cos(\tilde k_1\Delta t/2) \qquad \zeta_1= -\sin(\tilde k_1\Delta t/2)\Delta t/2\nonumber\\
\tilde k_1&=k_1+\nu_1 k_{g1}-\mu\omega_g
\end{align}
We use
\begin{align}
s_{l,i}&=\biggl(\frac{\sin(k_i\Delta x_i/2)}{\Delta x_i/2}\biggr)^{l+1}
\end{align}
as well as use the corresponding interpolation functions for the EM fields used to push the particles 
\begin{align}
S_{E1}&=s_{l,1}s_{l,2}s_{l,3}(-1)^{\nu_1}\qquad S_{E2}=s_{l,1}s_{l,2}s_{l,3} \qquad S_{E1}=s_{l,1}s_{l,2}s_{l,3}\nonumber\\
S_{B1}&=s_{l,1}s_{l,2}s_{l,3}\qquad S_{B2}=s_{l,1}s_{l,2}s_{l,3}(-1)^{\nu_1} \qquad S_{B3}=s_{l,1}s_{l,2}s_{l,3}(-1)^{\nu_1}
\end{align}
when using the momentum conserving field interpolation, and use
\begin{align}
S_{E1}&=s_{l-1,1}s_{l,2}s_{l,3}(-1)^{\nu_1}\qquad S_{E2}=s_{l,1}s_{l-1,2}s_{l,3} \qquad S_{E1}=s_{l,1}s_{l,2}s_{l-1,3}\nonumber\\
S_{B1}&=s_{l,1}s_{l-1,2}s_{l-1,3}\qquad S_{B2}=s_{l-1,1}s_{l,2}s_{l-1,3}(-1)^{\nu_1} \qquad S_{B3}=s_{l-1,1}s_{l-1,2}s_{l,3}(-1)^{\nu_1}
\end{align}
when using the energy conserving field interpolation. The $(-1)^{\nu_1}$ term is due to the half-grid offsets of these quantities in the $\hat 1$ direction. With respect to the current interpolation, \begin{align}
S_{j1} = s_{l-1,1}s_{l,2}s_{l,3}(-1)^{\nu_1}\qquad S_{j2} = s_{l,1}s_{l-1,2}s_{l,3}\qquad S_{j3} = s_{l,1}s_{l,2}s_{l-1,3}
\end{align} 
We note that we use expressions for charge conserving current deposition scheme that are strictly true in the limit of vanishing time step $\Delta t\rightarrow 0$. The coefficients $A_2$ to $D_2$ are real, and completely determined by $k_1$ and $k_2$. By solving Eq. (\ref{eq:asym1}) one can rapidly scan the NCI modes for a particular set of $(\mu,\nu_1)$.
\end{appendix}


\end{document}